\pdfoutput=1

\documentclass[acmsmall]{acmart} 
\AtBeginDocument{%
  }

\setcopyright{cc}
\setcctype{by}
\acmDOI{10.1145/3729400}
\acmYear{2025}
\acmJournal{PACMSE}
\acmVolume{2}
\acmNumber{FSE}
\acmArticle{FSE130}
\acmMonth{7}
\received{2024-09-12}
\received[accepted]{2025-04-01}

\acmConference[FSE '25]{ACM on Software Engineering, Volume 2, Number FSE}{June 23–27, 2025}{Trondheim, Norway}

\acmISBN{2994-970X/2025/7}

\usepackage{amsmath,amsfonts}
\usepackage{algorithmic}
\usepackage{graphicx}
\usepackage{xcolor}
\usepackage{hyperref}
\usepackage{cleveref}
\usepackage{xspace}
\usepackage{paralist}
\usepackage{multirow}
\usepackage{subcaption}
\usepackage{tikz}
\usepackage{listings}
\usepackage{booktabs}
\usepackage{wrapfig}
\usepackage{array}
\usepackage{fancyhdr}

\newcommand{\etal}{\emph{et al.}\xspace}

\newcommand{\eg}{\emph{e.g.},\xspace}

\usepackage{titlesec}
\definecolor{Gray}{gray}{0.3}
\tikzstyle{mybox} = [draw=black, very thick, rectangle, rounded corners, inner ysep=5pt, inner xsep=5pt, fill=gray!20]
\newcommand{\xyz}[2]{
    \smallskip
    \noindent
    \begin{tikzpicture}
        \node [mybox] (box){%
        \centering
        \begin{minipage}{.97\columnwidth}
        \fontsize{8.8}{10}\selectfont
        \textbf{RQ #1}. #2
        \end{minipage}
        };
    \end{tikzpicture}%
}

\begin{document}

\title{Calibration of Large Language Models on Code Summarization}


\author{Yuvraj Virk}
\affiliation{%
 \institution{UC Davis}
 \city{Davis}
 \state{California}
 \country{United States}}
\email{ysvirk@ucdavis.edu}

\author{Premkumar Devanbu}
\affiliation{%
 \institution{UC Davis}
 \city{Davis}
 \state{California}
 \country{United States}}
\email{ptdevanbu@ucdavis.edu}

\author{Toufique Ahmed}
\authornote{Majority of the work was done when the author was a postdoctoral scholar at UC Davis.}
\affiliation{%
 \institution{UC Davis \& IBM Research}
 \city{Yorktown Heights}
 \state{New York}
 \country{United States}}
\email{tfahmed@ibm.com}

\begin{abstract}
A brief, fluent, and relevant summary can be helpful during program comprehension; however, such a summary does require significant human effort to produce. Often, good summaries are unavailable in software projects, which makes maintenance more difficult. There has been a considerable body of research into automated AI-based methods, using Large Language models (LLMs), to generate summaries of code; there also has been quite a bit of work on ways to measure the performance of such summarization methods, with special attention paid to how closely these AI-generated summaries resemble a summary a human might have produced. Measures such as BERTScore and BLEU have been suggested and evaluated with human-subject studies.  

However, LLM-generated summaries can be inaccurate, incomplete, \emph{etc} : generally, too dissimilar to one that a good developer might write. Given an LLM-generated code summary, how can a user rationally judge if a summary is sufficiently good and reliable? 
Given just some input source code, and an LLM-generated summary, 
existing approaches can help judge brevity, fluency and relevance of the summary; 
however, it's difficult
to gauge whether an LLM-generated summary
sufficiently resembles what a human might produce, without a ``golden" human-produced summary
to compare against. 
We study this resemblance question as a \emph{calibration} problem: given just the code \& the summary from an LLM, can we compute a \emph{confidence} measure, that provides a reliable indication of whether the summary sufficiently resembles what a human would have produced in this situation? We examine this question using several LLMs, for several languages, and in several different settings. Our investigation suggests approaches to provide reliable predictions of the likelihood that
an LLM-generated summary would sufficiently resemble a summary a human might write for the same code. 
\end{abstract}





\maketitle
\section{Introduction}
\label{intro}
Code summaries and comments help maintainers understand code. \cite{hu2022practitioners} surveyed developers in several major software development organizations on software summarization practices. The survey responses indicated that developers value code summaries and, furthermore, that they would appreciate tools to automatically generate such summaries for them. Automated Code Summarization has been a very active area of research~\cite{zhang2022survey, zhu2019automatic} for quite a while. Deep learning has helped advance summarization performance to new heights, with the current state-of-the-art achieved via the use of in-context learning (or prompt engineering) with large language models (LLM)~\cite{ahmed2024automatic, nashid2023retrieval}. 

\begin{figure}[t]
  \centering
  \begin{minipage}[t]{0.48\linewidth}  
    \vtop{
      \vskip 1pt
      \hbox{
        \includegraphics[width=\linewidth]{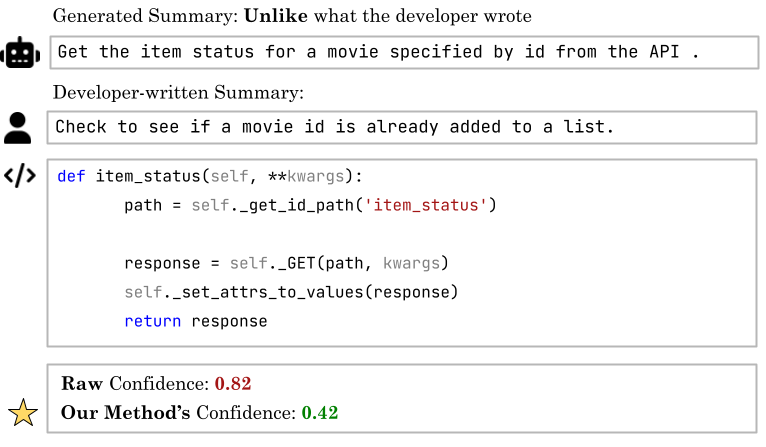}
      }
    } 
  \end{minipage}
  \begin{minipage}[t]{0.48\linewidth}  
    \vtop{
      \vskip 0pt
      \hbox{
        \includegraphics[width=\linewidth]{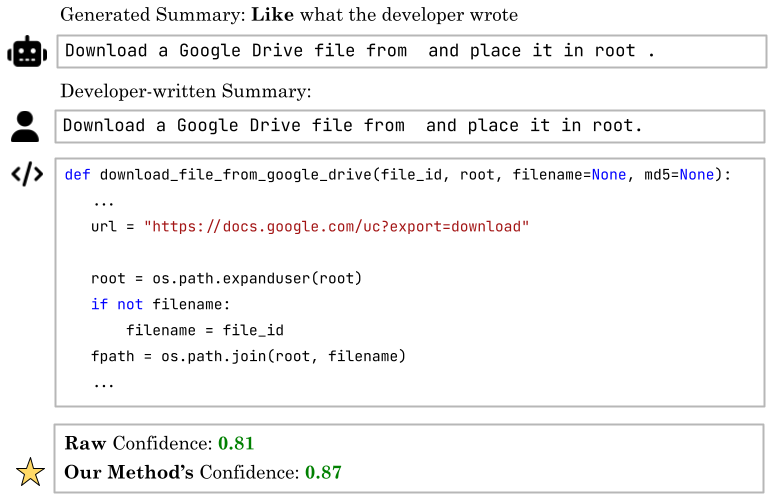}
      }
    }
  \end{minipage}
  \vspace{-.1cm}  
  \caption{Example of two LLM-generated summaries with nearly identical confidences but differing quality. The generated summary on the left is dissimilar to what the developer wrote, but the confidence is high. The generated summary on the right is similar to what the developer wrote, and the confidence is high, as it should be; however, the confidence is also high on the left! Our method preserves the high confidence for the good summary but produces a low confidence for the poor one.}
  \label{examp-summary}
  \vspace{-.4cm}  
\end{figure}

Although LLMs are increasingly effective at code summarization, sometimes even surpassing the quality of human-written summaries (discussed in Section \ref{discussion}), they still have lapses in quality \cite{su2024distillgptcodesummarization,afrin2025resourceefficienteffectivecode}. For example, 46.3\% of the time, humans judge LLM-generated code summaries as worse than developer-written ones, at explaining the role of a method \cite{su2024contextawarecodesummarygeneration}. In our study, 70-80\% of the time, LLMs produce summaries humans may not find acceptably similar to human summaries. As LLMs generate code summaries at scale, inaccurate or incomplete code summaries even 1\% of the time may significantly hurt program comprehension if relied upon.

Thus, given an AI-generated summary, it would be helpful to have an indication as to whether that summary is an “acceptable" one. For reasons clarified below in Section~\ref{sec:bg}, and due to the importance that prior work places on it~\cite{hu2022practitioners, shi2022evaluation, mastropaolo2024evaluating}, we use \emph{``acceptable" similarity to human-written summaries} as our primary correctness measure. When used together other metrics, this measure helps holistically evaluate a code summary's quality.

But how do we gauge whether the summary will be acceptable or not without the presence of the reference? In NLP~\cite{guo2017calibration}, model log-probabilities have been widely studied as confidence measures for classification tasks, with various scaling methods proposed to improve their reliability. More recently, similar techniques have been applied to assess LLM calibration in SE tasks~\cite{lookleap}, including program repair~\cite{zhou2024calibration,zhang2023pace,spiess2024calibration}. Can the raw model confidence, directly derived from the model’s output, reliably indicate whether the summary will be acceptable, without access to a reference?

If this confidence measure is reliable, then a summary produced with high confidence can be accepted verbatim and used directly to help understand the code; summaries at low confidence might be simply rejected; and summaries at a medium confidence might be considered partially useful, requiring some further examination of the code. This is where calibration (discussed in Section~\ref{sec:bg}) becomes relevant. Calibration refers to the agreement between the predicted confidence of an event and the actual event outcome. A well-calibrated model will produce probability estimates that match the true likelihood of an event. For example, if a model predicts that an event has a 80\% chance of occurring, then that event should indeed occur approximately 80\% of the time across many predictions with the same probability. The model's high confidence in a low-quality summary in Figure~\ref{examp-summary} illustrates how LLMs can be poorly-calibrated for code summarization.  

In this paper, we study the issue of providing a  confidence measure, that is a reliable indicator of how likely a  generated summary is correct, \emph{viz.,} sufficiently similar to a human-written summary. For instance, in Figure~\ref{examp-summary}, had our approach been used, then the more desirable outcome in fact comes with a higher confidence (0.87 vs 0.42); thus, our approach would support better decision-making regarding the reliability of a summary. 

We make the following contributions.
\begin{itemize}
\item We introduce the problem of calibration for code summarizers; we frame ``correctness" of generated summaries in terms of being \emph{sufficiently similar} to human-generated summaries. 
\item Using several thresholds of similarity measures, corresponding to several settings of ``correctness", we examine the calibration of LLMs. 
\item 
We show how LLMs augmented with in-context learning, and using
modern rescaling techniques, can provide better calibration for the code summarization task. 
\item 
While later tokens in a long code summary have very high probabilities and exhibit overconfidence, earlier tokens are better calibrated.
\end{itemize}

Our findings provide a way for developers to make well-justified decisions regarding the use of (sometimes low-quality) summaries generated by LLMs. 

\section{Background \& Motivation}

\label{sec:bg}

We begin with a brief summary of the current  state of code summarization, and immediately dive into the problem of calibration. 

Code summarization is important for code understanding and maintenance~\cite{sridhara2010towards}; 
building tools that can automatically summarize a given piece of code (method
or snippet) is an important challenge that has received considerable attention~\cite{zhu2019automatic,zhang2022survey}. 
Neural approaches have been applied to code summarization. Earlier 
approaches used \emph{pre-training + fine-tuning}, with models such as CodeBERT~\cite{feng2020codebert}, PLBART~\cite{ahmad2021unified}, PolyGlot~\cite{ahmed2022multilingual}, and CodeT5~\cite{wang2021codet5}.
Current approaches exploit Large Language Models (LLMs) ~\cite{brown2020language,nashid2023retrieval, su2024distillgptcodesummarization, afrin2025resourceefficienteffectivecode, su2024contextawarecodesummarygeneration}; 
with such powerful LLMs, \emph{in-context learning} techniques \emph{e.g} few-shotting~\cite{brown2020language,ahmed2022few}, chain-of-thought \cite{sun2024sourcecodesummarizationera}, and semantic augmentation~\cite{ahmed2024automatic}, attain state-of-the art code summarization performance without any additional training. 
In this paper, our focus is on LLM-based code summarizers that use in-context learning and semantic augmentation. Despite the progress in code summarization, code summaries are still produced by imperfect models.

\subsection{Calibration: How to use unreliable outputs from flawed models}
Neural models are probabilistically optimized over the training set. On a distinct test set, they may not always produce high-quality results. How then, is one to use these
flawed models in practice? 
This question relates to the concept of \emph{Calibration}.

For simplicity, let us first consider
a neural model that produces a simple binary prediction, which is not always correct. Let's further assume that the prediction is associated with a confidence, which is a probability measure ($0 <= p <= 1$). For this prediction to be reliable, we would like the output \emph{confidence} to reliably indicate
its empirical rate of \emph{correctness}; thus whenever the model is 60\% confident (\emph{i.e,} output probability is 0.6) then we would like to be correct about 60\% of the time.  If this is the case, the model
is said to be \emph{well-calibrated}; model confidence is a good predictor of expected outcomes,
and thus helpful in determining the expected value of a decision policy. 
With a well-calibrated weather-forecasting model, \emph{e.g},  users can respond appropriately to different confidence levels: for instance, using a hat at 10\% confidence of rain, an umbrella at 50\% or above, or taking a bus at 80\% or above. 

\begin{wrapfigure}{r}{0.53\linewidth}
\vspace{-10pt}
  \centering
  \begin{minipage}[b]{0.45\linewidth}
    \includegraphics[width=\linewidth]{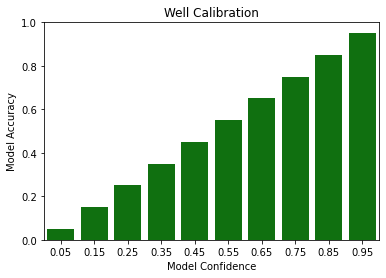}
  \end{minipage}
  \begin{minipage}[b]{0.450\linewidth} 
    \includegraphics[width=\linewidth]{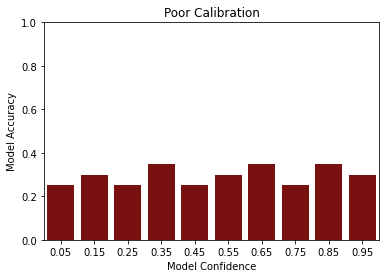} 
  \end{minipage}

  \caption{Sample of well and poorly calibrated models}
  \vspace{-10pt}
  \label{examp-calib}
\end{wrapfigure}
Quantitatively, \emph{Calibration} is a relationship (Figure \ref{examp-calib}) between two quantities: \emph{confidence} on the x-axis, typically a probability value, and \emph{the rate of correctness} (y-axis), typically a normalized frequency (values  $ \in [0,1]$) indicating how often the model is \emph{empirically
correct}, when predicting at the confidence values along the x-axis. 
Figure~\ref{examp-calib} presents two \emph{reliability plots}~\cite{wilks1990weather} to show  the difference between a well-calibrated model and a poorly calibrated model; in the plot, the predicted confidences are binned along the x-axis, over 10 ranges $0-0.1,0.1-0.2,\ldots$, and the fraction of each bin that is correctly predicted is shown on the y-axis. In a well-calibrated model, the observed model correctness rate is well-aligned with the model confidence; for a poorly calibrated model, it is not. Clearly, improving
calibration is a worthy goal; to do so, we first have to measure it. 

\subsubsection{Measuring Calibration.} 
A  \emph{calibration} metric measures the alignment of a model's \emph{confidence} (in its predictions)  with the \emph{correctness} thereof.
We now discuss two main measures, Brier score, and ECE. Note again, we first consider models with binary
outcomes: either the model prediction is correct, or it is wrong. 

\vspace{.1cm}

\noindent\textbf{Brier Score}\footnote{See  \cite{brier1950verification} for details.} is the mean squared difference between the predicted probabilities and the actual outcomes; it measures (on average)
how well (given a sample) predicted confidence aligns with the actual correctness on that sample. 
A perfectly calibrated model would have a Brier Score of 0, indicating that its predicted probabilities match the observed outcomes exactly. Mathematically, the Brier Score $B_r$ 
for a sample set of $N$ predictions is calculated as follows:

\vspace{-0.12in}
\begin{equation}
B_r = \frac{1}{N} \sum_{i=1}^{N} (\hat{p_i} - \hat{o_i})^2
\label{brier_equation}
\end{equation}
\vspace{-0.05in}

\vspace{-0.01in}

\noindent Where $\hat{p_i}$ is probability (confidence) assigned to the $i^{th}$ sample prediction, and $\hat{o_i}$ is the $i^{th}$ sample outcome (1 for correct, 0 for incorrect).
Lower $B_r$ indicate better calibration, with 0 being perfect calibration, with a maximum (worst) possible score of 1.
$B_r$ thus measures the alignment of a model's confidence estimates with the actual rate of empirical correctness.

Brier scores should always be gauged relative to a \emph{base rate}. 
A na\"{i}ve model, which always outputs (as confidence) the empirical base rate $p$, 
(for example, if it rains 10\% of the days, the model predicts rain with $p=0.1$ confidence every day) 
will achieve a score
called the \emph{reference Brier score}, $B_{ref}$, which analytically is $p(1-p)$.
Thus, if it rains about 10\% of the days, the base rate is $0.1$; a model that always outputs
 $0.1$ as its confidence will score $0.1*0.9 = 0.09$ Brier. Evidently,
$B_{ref} \in (0,0.25]$.  
 
Given the possibility of  low Brier scores from actually very ``unskilled" na\"{i}ve models, we need to normalize the measured Brier for a given model; we discuss this below, as ``skill scores". 

\vspace{0.1cm}

\noindent\textbf{Brier Skill Score}  is a normalized measure $\in (-\infty,1] $
for assessing the reliability of a model's confidence, 
relative to a  na\"{i}ve model always just providing the base rate as the confidence,
and a skill of $B_{ref}$.
A better-calibrated model will achieve a Brier Score lower than this unskilled $B_{ref}$ value; a bad model,
could do worse! The Skill Score (SS) quantifies this improvement
(or decline) compared to the baseline $B_{ref}$, calculated as: 
\begin{equation}
SS = \frac{(B_{ref} - B_{model})}{B_{ref}} 
\label{skill_score_eq}
\end{equation}
A positive SS 
indicates improvement over the baseline, while a negative SS suggests predictions worse than the baseline. A perfect skill score is 1.0; but even small positive values of SS can sometimes indicate good skill.
The German weather forecasting service, \emph{Deutsche Wetterdienst}, sets a minimum threshold of 0.05 for a Skill Score to indicate good forecast quality reference~\cite{WetterUndKlimaa}. Another reference point comes from the American data journalism site \emph{538}, which reports a skill score of approximately 0.13 in forecasting World Cup games~\cite{wezerekHowGoodAre2019}.

However, Brier isn't the only metric that's used. 
\vspace{.1cm}

\noindent\textbf{ECE.}
(Expected Calibration Error)~\cite{naeini2015obtaining} is another,
rather more complex, calibration measure. It averages the difference between the predicted confidence and actual correctness rates across different confidence levels. 
Intuitively, on a reliability diagram (such as \autoref{examp-calib}) it amounts to measuring
the (normalized) area between the diagonal ideal and the actual empirical bars. 
A lower ECE indicates better calibration, meaning the predicted probabilities closely match the actual rates. Conversely, a higher ECE indicates poorer calibration, meaning a mismatch between predicted and actual rates.

\begin{wrapfigure}{r}{0.5\textwidth}
\vspace{-12px}
\begin{minipage}{0.5\textwidth}
\begin{equation}
acc(B_i) = \frac{1}{|B_i|} \sum_{\hat{o_i}\in B_i} {\hat{o_i}}
\end{equation}

\begin{equation}
conf(B_i) = \frac{1}{|B_i|} \sum_{\hat{p_i}\in B_i} {\hat{p_i}}
\end{equation}

\begin{equation}
	ECE = \displaystyle\sum_{i=1}^m \frac{\mid B_i \mid}{\mid T \mid} \lvert \text{acc}(B_i) - \text{conf}(B_i) \rvert
\end{equation}
\vspace{-8px}
\end{minipage}
\end{wrapfigure}

To compute ECE, we first partition the dataset into $n$ bins $B_i, i = 1 \ldots n$ based on confidences (predicted probabilities). For example, the bins could range from confidence levels of [0-0.1) to [0.9-1.0). For each bin, we calculate the average confidence, $conf(B_i)$ (predicted probability), and empirical accuracy, $acc(B_i)$ (fraction of true positives). Following this, we determined a weighted difference by (i) computing the absolute difference between the average confidence and accuracy in each bin and (ii) weighting by the proportion of samples in each bin. Finally, we sum the weighted differences across all bins to get the ECE. 
ECE provides a more nuanced evaluation of calibration across different confidence levels compared to metrics like the Brier Score, which aggregates errors across all samples without considering confidence levels. This makes ECE particularly useful for understanding how well a model's confidence estimates align with its performance at different confidence levels. Mathematically, the ECE is calculated in three steps below (\emph{Note}: $B_i$ denotes the set of elements in each bin; $\hat{o_i}$ is an indicator varible, $=1$
if the prediction is right, and $=0$ if wrong; $\hat{p_i}$ is the confidence (probability) associated with the prediction)

ECE can be intuitive, and visually appealing: \emph{but it can mislead}, since \emph{the binning
approach may vary}. 
For example, if it usually rains 10 days a year in Los Angeles, a na\"{i}ve (technically, ``unskilled") model could always predict rain with a probability of $\frac{10}{365}= 0.0275$. Such a model
would have a single bin with a confidence of $0.0275$, and an empirical correctness rate of nearly the same (on a yearly basis), yielding a perfect (but misleading) ECE of zero.

ECE and Brier Score serve slightly different purposes: the Brier Score measures the ability to correctly discriminate output categories and the calibration of output probability for each sample, while the ECE specifically measures calibration. However, the ECE can be misleadingly low, as noted earlier for the unskilled predictor. Additionally, careful binning is necessary as it can impact ECE scores.

Thus far, we have considered calibration with respect to a model with a binary output. What if the output is a sequence of tokens, like a code summary? Calibration in this situation is a bit more complex, but we still need 
two notions: \emph{confidence}, and \emph{correctness} which are used to calculate Brier Skill score, 
and $ECE$. We begin with
approaches to evaluate \emph{confidence} of model-generated code summaries, and then turn to \emph{correctness}.

\subsection{Confidence in a generated summary}
Thus far, our discussion of calibration has focused  settings where there is a \emph{single, binary} prediction, with an associated confidence (probability) measure. With generated code summaries, we have a sequence of tokens generated from a softmax layer; from this layer, we can sequentially select a most likely token, with an associated probability. This would be ``greedy" decoding; other methods are possible. Whatever method is used, one obtains a generated code summary, which contains a sequence of tokens, each with an associated probability. Since calibration measures rely on a single confidence associated with an entire output, we  need a way to summarize the entire sequence of probabilities. The simplest approach
is to just take a mean (geometric or arithmetic) of this sequence. 
We discuss further details in the methods section. 


\subsection{Correctness of a generated summary}
\label{sec:summaryeval}


For a binary prediction (\emph{e.g.} ``Will it rain tomorrow'') the notion of correctness
is simple (either it rained or didn't). For a generated sequence of tokens constituting a code summary, what exactly is correctness? Exact match? Partial match? Similarity of ``meaning''? 
Fortunately, there has been a quite a bit  of work on  evaluations of code
summaries~\cite{roy2021reassessing,shi2022evaluation,haque2022semantic,mastropaolo2024evaluating}. 

Prior work emphasizes evaluating code summaries by their similarity to human-written summaries \cite{haque2022semantic, shi2022evaluation, stapleton2020human}, in conjunction with other metrics like SIDE \cite{mastropaolo2024evaluating}. Human-written summaries capture content actual developers deemed useful for program comprehension; similarity to them thus serves as \emph{one} of multiple signals for summary quality. \cite{hu2022practitioners} conducted a human-subject study to identify the criteria 
that affect practitioners' judgment of a code summarization technique's performance. The important evaluation criteria include \emph{content adequacy},  \emph{conciseness}, \emph{fluency}, \emph{useful information content \underline{not} in code}, and \emph{similarity to original human-written comments}. It should be noted that of the above 5 criteria, the first 4 could potentially be evaluated using just the source code and generated summary; the last requires access to a human-written summary, which is typically not available when one is using a LLM to generate code summaries. In this paper, we therefore focus on \emph{similarity to human-written summary} as our ``goodness" criteria. 

Researchers have sought the right metrics to measure the similarity of a machine-generated summary to a human-produced summary ~\cite{haque2022semantic,roy2021reassessing,shi2022evaluation,stapleton2020human}. Metrics such as BLEU~\cite{papineni2002bleu}, ROGUE~\cite{lin2004rouge}, METEOR~\cite{banerjee2005meteor} have been used to measure lexical (word n-gram level) similarity; but these have been criticized as not being well-aligned to human judgements of similarity in the case of code summaries~\cite{stapleton2020human}. ~\cite{haque2022semantic}  empirically compared several different
lexical \& semantic metrics of similarity, and found that certain embedding-based metrics such as BERTScore are better-correlated with human evaluations of summary similarity (to a reference) than purely lexical measures. They report that SentenceBERT has the highest \emph{correlation} with human evaluations of summary similarity. 

We re-analyzed all the data \& metrics used by ~\cite{haque2022semantic}, from a somewhat different, binary decision
perspective: 
\emph{Can we reliably predict {\bf sufficient} {\bf similarity}, of a model-generated summary}? 
For our perspective we found that both BERTScore and SentenceBERT could be useful. 
Note that in our setting, we seek to \emph{predict} sufficiently high semantic similarity (above a threshold value, as discussed in ~\autoref{subsec:thresholding}) of a generated summary, without knowing the Gold (human-generated) summary. 
We would like to generate a well-calibrated confidence signal (associated
with the generated summary) that is a \emph{reliable} indicator
of the empirical likelihood of sufficient semantic similarity between generated summary and the gold summary (were it available) being high enough. Our re-analysis of the
Haque et al data is discussed in methods section, \S~\ref{subsec:thresholding}. 

\vspace{0.05in}
\noindent{\bf An Example Scenario:} To illustrate our approach: consider a user $\mathcal{U}$ who is dealing with code $C$
that has no human-written summary; $\mathcal{U}$ then generates a summary $S$ from $C$, using a neural model.
How would $\mathcal{U}$ evaluate $S$, relative to the code $C$? Considering Hu \emph{et al}'s 5 criteria: $\mathcal{U}$ could judge \emph{fluency} and \emph{conciseness} just from reading $S$. Recently, ~\cite{mastropaolo2024evaluating} introduced SIDE, a way to automatically measure the alignment of summary $S$ to $C$ using just the code and the candidate summary (and no human-written reference); it was found to strongly predict human ratings of content adequacy. Finally, turning to \emph{similarity to developer-written comments}, $\mathcal{U}$ currently has no way develop a reliable level of confidence in $S$, without access to the developer-written summary. Given the importance prior work places on similarity to human-written summary, we focus our attention on developing a well-calibrated confidence predictive of such similarity. Finally, regarding \emph{useful content not in the code}, $\mathcal{U}$ would have to read $C$ and $S$, or potentially use our confidence measure as a proxy; developer-written comments we compare to can incorporate working knowledge of the software process outside the code. Now, provided with a well-calibrated confidence for $S$, $\mathcal{U}$ could make a decision that best suits their needs. Thus, if $C$ is critical for $\mathcal{U}$'s work, and is really important to understand,  $\mathcal{U}$ might accept only very high-confidence summaries, and ask a human in all other cases. In cases where $C$ is less critical, and a surface understanding is sufficient, a moderate level of confidence in $S$ might be sufficient. If $S$ is delivered at a very low confidence, $\mathcal{U}$ might just reject it. 

The above scenario illustrates the value of associating a well-calibrated confidence with a \underline{\emph{summary}}
generated by a neural model. A different, but analogous scenario, regarding the use of  well-calibrated confidences 
by a user considering whether to use model-generated \underline{\emph{code}},
has been discussed in prior work~\cite{lookleap,rusure,spiess2024calibration}. We are
now ready to present our research questions. 

\section{Research Questions}
\label{rqs}
Given an input code $C$ and a generated summary $S$, 
our  study aims to find a way to compute well-calibrated \emph{confidence} signal, associated
with a useful standard of \emph{correctness} for $S$. Our research questions are animated
by this central goal. 

In the first research question, we study whether an LLM's confidence in its generated summary can be used to predict the quality of its generated summary. As previously established, we evaluate summary quality using metrics measuring BERTScore similarity to human-produced summaries. 
We first explore whether an LLM's log-probabilities can be directly used as a reliable \emph{confidence} signal
of the empirical likelihood of such similarity. For \emph{correctness}, 
we transform the continuous BERTScore value into a binary indication of
sufficient similarity, using thresholding: a summary is considered correct if
it has a high enough BERTScore value. 
We try several approaches to obtaining a summary confidence measure
from per-token LLM-generated probabilities, and several approaches to BERTScore
thresholding. These are
further discussed in Methods Section~\ref{thresholding}.


\xyz {1} {Are
the LLM-produced confidence measures over  the tokens in the generated summary well-calibrated?}

We find that confidence measures directly calculated by averaging LLM per-token probabilities are not well-calibrated. We try 
\emph{Platt rescaling}, which is a method used to adjust a model's confidences to better align with the observed frequencies of events. 

\xyz {2} {How does rescaling affect the calibration of various LLMs with respect to summary correctness?}

In the first two research questions, we measured a LLM's confidence using its own calculated probability (confidence),
over \emph{all} the tokens in the generated output. However, in our examination of the calculated probabilities
from the model, we found that per-token probabilities change quite a bit from the beginning of
the generated sequence, towards the later tokens; we studied potential differences in calibration 
across the length of the generated sequence. 

\xyz {3} {Are earlier tokens in the generated summary better calibrated than later tokens?}

These questions taken together, constitute, to our knowledge, the first detailed examination of the calibration of LLMs with respect to the code summarization task. 

\section{Methodology}
\label{sec:methods}
\label{sec:data}
We describe the  models \& prompting strategies we used for the experiments; 
and then get into our approach to calculating
confidence,  determining correctness, and calculating calibration. 

\subsection{Models}

\vspace{.1cm}

\noindent\textbf{GPT-3.5-Turbo.}
GPT-3.5 models perform well at both comprehension and generation, for both natural language and code~\cite{guo2024deepseek}. At the time of experimentation, GPT-3.5-Turbo was the most capable and affordable option among these models. Although it is primarily optimized for chat-based interactions, it also performs well in non-chat tasks. GPT-3.5-Turbo\footnote{https://platform.openai.com/docs/models/gpt-3-5-turbo} can generate up to 16,385 tokens, including the prompt tokens.

\vspace{.1cm}

\noindent\textbf{Code-Llama-70b.}
The Code Llama~\cite{roziere2023code},  family of  language models are tailored for coding tasks. It offers 
leading performance among open models. It has infilling capabilities, support for extensive input contexts, and zero-shot instruction-following abilities. The family includes foundation models, Python-specialized variants, and instruction-following models, each with 7B, 13B, 34B, and 70B parameters. Trained on 16k-token sequences, these models support  inputs of up to 100k tokens. Notably, Code Llama achieved state-of-the-art results on various code benchmarks at the time of experimentation. We use the larger 70B parameter model for our experiments.

\vspace{.1cm}

\noindent\textbf{DeepSeek-Coder-33b Instruct.}
The DeepSeek-Coder series introduces open-source code models ranging from 1.3B to 33B parameters, trained from scratch on a 2 trillion-token dataset~\cite{guo2024deepseek}. These models, pre-trained on high-quality project-level code corpora, utilize a fill-in-the-blank task with a 16K window to enhance code generation and infilling. Extensive evaluations demonstrate that DeepSeek-Coder achieves state-of-the-art performance among open-source code models across various benchmarks, outperforming closed-source models like Codex and GPT-3.5. Our study uses the 33B paramater model.

\subsection{Prompting Methods for Code Summaries}
\label{csmethod}
The above models are used with state-of-the-art prompting techniques that have been demonstrated to work well for code summarization. 

\vspace{.1cm}

\noindent\textbf{Retrieval Augmented Few-shot Learning.}
Few-shot learning works well for both Natural Language Processing~\cite{brown2020language} and Software Engineering~\cite{ahmed2022few,nashid2023retrieval}: we prompt models with 3 exemplars as query-answer pairs (method-comment pairs) and instruct the model to answer our final query (just the method).
Prior works show that few-shot learning performance can be improved using exemplars \emph{relevant} to the final query,  identified using a retrieval algorithm~\cite{ahmed2024automatic,nashid2023retrieval}. 
Prior work also shows that the BM25 retrieval algorithm 
works better than vanilla few-shot learning for many SE tasks (\eg code summarization, program repair, and assertion generation), so in this paper, 
we use few-shots retrieved with BM25. 

\vspace{.1cm}

\noindent\textbf{Automatic Semantic Augmentation of Prompt (ASAP).}
Chain-of-thought~\cite{wei2022chain} has been found to be effective for several tasks in NLP: providing the model with examples comprising explicit intermediate reasoning steps improves performance. 
For SE, 
\cite{ahmed2024automatic} showed a workable approach (ASAP) to 
extracting such ``intermediate steps" using static analysis algorithms; this approach  enhances model performance. They used GitHub repository information, identifier type with scope, and dataflow information. Combined with BM25, this approach yielded state-of-the-art performance for code summarization tasks~\cite{ahmed2024automatic}. We therefore also use BM25 few-shot together with ASAP-based chain-of-thought. 

\subsection{How to measure models' confidence in a token sequence? (RQ1)} 
\label{model_confidence}

An LLM’s per-token probabilities reflect the likelihood of a token appearing in the same context in the (human-produced) training data.
Therefore, we may expect a sequence-level probability to reflect the likelihood a generated summary is similar to a human-produced one. Still, the probabilities
are calculated by the model per-token; so given a sequence of $N$ per-token probabilities, how should we compute a single representative probability?

The sequence-level probability (under an independence assumption) is the product of per-token probabilities. Since the product is usually very small and varies with sequence length, we normalize by taking its $N$-th root \emph{i.e.} the geometric mean of probabilities. 
The arithmetic mean offers another option; however, 
since the product, not the sum, of individual probabilities better represents the probability of a sequence, we prefer the geometric mean. Empirically, we also consistently found better calibration when using the geometric mean over the arithmetic mean, since it less sensitive to large token-probability values (and thus reduces risk of overconfidence).

\subsection{Evaluating similarity to human-produced summaries (RQ1)}
\label{subsec:thresholding}
In Section \ref{sec:summaryeval} we described the range of available metrics, and justified our focus on measures of the similarity to human-produced summaries. To evaluate the calibration of code summarization, we need a binary notion of correctness which separates a good and bad summary.
\label{thresholding}
Our notion of correctness is \emph{sufficient similarity} to human-produced summaries; this requires
thresholding similarity metrics, which are real-valued quantities. 
We first determine these thresholds using available data  an existing human subject studies, and then use these thresholds to study calibration performance of 3 models, 
on a much larger dataset used
for studies of automated code summarization. 
\vspace{.1cm}

\noindent\textbf{Dataset used for Setting Correctness Threshold }
We use the ~\cite{haque2022semantic} dataset, which includes reference and generated summaries for 210 Java methods. For each Java method, it includes 3 human-assigned similarity ratings. Human raters select from 4 ratings ranging from "Strongly Disagree" to "Strongly Agree" indicating their agreement that the generated summary is similar to the reference. ~\cite{haque2022semantic} used their data to analyze several similarity metrics, and found that the semantic metric SentenceBERT offered the best \emph{correlation} with human ratings. 

Following their methodology, for each code method, we take the mean of the 3 human ratings to produce an aggregate rating. We use this dataset to evaluate how accurately different automatic similarity metrics classify human agreement vs. disagreement on a summary's similarity to the reference summary. For our purposes,
to qualify as a ``correct" summary, the average human rating in the dataset must indicate at least an \underline{``Agree"} rating. 

\vspace{0.1in}
\noindent\textbf{Methodology for Selecting Thresholds}
We wanted thresholds for automated similarity metrics that are most consistent with this ``Agree" 
level of human judgement of similarity. 
\begin{wrapfigure}{r}{0.5\linewidth} 
  \vspace{-15pt} 
  \centering
  \includegraphics[width=\linewidth]{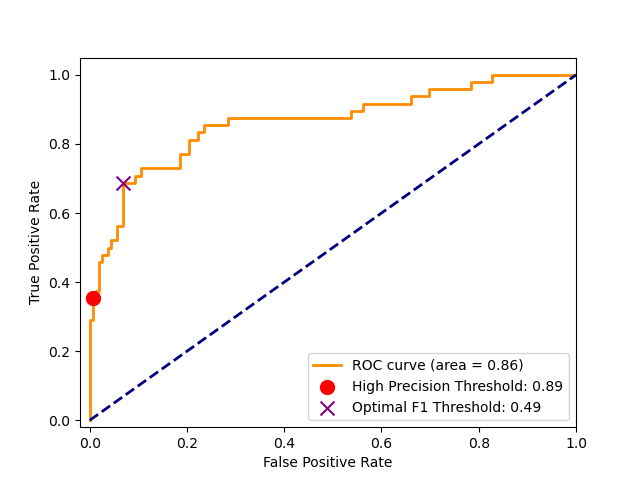}
  \vspace{-20pt}
  \caption{BERT-Score ROC Curve}
  \label{BERT-Score-ROC-Curve}
  \vspace{-10pt} 
\end{wrapfigure}
An optimal metric and threshold should classify all summaries that humans judge as similar as correct (perfect recall) while avoiding wrong classifications (perfect precision). As expected, all the metrics in Haque \etal were imperfect; however, for illustration, with our criterion for correctness, we show an AUC-ROC curve for BERTScore  (\autoref{BERT-Score-ROC-Curve}) with an AUC of 0.86, which suggested that this metric, while not
perfect, was a reasonable proxy for human judgement of similarity.

Given imperfect metrics, there are different ways to select thresholds, based on the desired outcome.  Lower thresholds would offer higher recall, but risk including summaries that may not resemble gold summaries. Higher thresholds would include fewer summaries, but ones more likely to resemble the gold ones. A cautious developer may want higher precision, lower recall; a less cautious developer might want some reasonable summary, even if lacking assurance of similarity to human summaries. For our purposes here, we would like to evaluate calibration for a set of thresholds over several different trade-offs, specifically: high recall,
high precision, and  optimal F1.

\indent To identify workable thresholds, we perform a grid search over the similarity metrics  (from Haque \etal), and threshold values, at 0.01 intervals to identify metric-threshold pairs meeting
several criteria: 1) the highest recall while maintaining a high precision (> 0.9), 2) the highest precision with high recall (> 0.9), 3) the highest F1 score (harmonic mean of precision and recall). These metrics and thresholds respectively prioritize cautious, lenient, and balanced classifications of good vs. bad summaries, as viewed from the human similarity judgment data. For each condition, the grid search produces: 1) BERTScore at threshold 0.89 for high precision 2) Sentence-Bert at threshold 0.80 for high recall and 
3) BERTScore at threshold 0.49 for optimal F1. We report results for all
these settings. For illustration, Figure~\ref{BERT-Score-ROC-Curve} shows the two threshold values selected for  BERTScore.  

\subsection{Experimental Dataset}
\label{sec:dataset}
After setting the correctness thresholds for BERTScore
using the dataset from~\cite{haque2022semantic},  we needed a larger dataset for the calibration experiments. 
We use the Java and Python code summarization datasets from the CodeXGLUE Benchmark ~\cite{lu2021codexglue} for  evaluating the calibration of
the confidence values generated from LLM per-token probabilities. 
CodeXGLUE includes 14 datasets covering 10 Software Engineering tasks. Java and Python datasets contain 164,923 and 251,820 training samples, respectively. We use BM25~\cite{robertson2009probabilistic} algorithms to retrieve relevant samples from these datasets for few-shotting. For the test set, we randomly select 5000 samples from each language. Only the training partition was 
used for BM25 retrieval. 
It should be noted that the CodeXGLUE has been de-duplicated, reducing the risk that an example from
the test set would appear as a retrieved few-shot sample. It should also be noted
that reference (human-written) summaries are available in this dataset; generated
summaries can be compared with reference summaries to see if the semantic similarity score
is above the given threshold. 
While some human references in this dataset collected from open-source may be of poor quality \cite{commentsevolution}, the dataset as a whole captures content deemed as useful by actual developers. Among current evaluation approaches, the similarity to these docstrings captures \emph{this} aspect best; casual human raters may lack the specific and deep knowledge the original developers have.

\subsection{Rescaling (RQ2)}
\label{rescal}
Rescaling involves adjusting the predicted probabilities of a model to better align with the observed frequencies of events.
Platt Scaling~\cite{platt1999probabilistic} is a commonly employed method in calibration, where a logistic regression model is fitted to the predicted probabilities alongside the corresponding true outcomes. Through this process, the model transforms the predicted probabilities to more accurately reflect the true probabilities of events. While there are other rescaling approaches 
~\cite{guo2017calibration, zadrozny2001obtaining,zadrozny2002transforming}, we
use Platt scaling, since it has shown success in prior software engineering work~\cite{spiess2024calibration}. 

To fit the Platt rescaler, we employ 5-fold cross-validation, where for each fold, we utilize the other 4 folds as training data and apply the learned regression to our target fold. We repeat this process for all other folds to obtain rescaled values for all samples. To accurately evaluate that rescaling is robust, each fold only contains samples from different repositories. This way, test samples are always from different repositories than training samples; the reported Brier and Skill scores are calculated cumulatively over all folds. 

\subsection{Using the first tokens to compute confidences (RQ3)}
\label{token_cutoff}
Each token generated in a summary sequence successively constrains the next tokens in the sequence. For instance, the first few tokens might include an identifier, making the next verbs, prepositions, nouns, \emph{etc.} most related to that identifier. Our initial study suggested that later tokens in a code summary have higher probabilities. 

Thus, we expect that an initial token-sequence of length $t$ might
provide the best-calibrated summary confidence metric; this $t$  becomes a hyper-parameter to be tuned. 
Our method selects the parameter $t$ that maximizes skill score. Similar to Platt Scaling, we employ 5-fold cross-validation. For each fold, we utilize the other 4 folds to identify the number of tokens $t$ that maximizes skill score, and apply this hyperparameter $t$ to our target fold. We repeat this process for all other folds to obtain calibration results using the first $t$ tokens for all samples.

In order to test whether using the first $t$ tokens significantly improves calibration over using all tokens, we test the statistical significance of the difference in Brier scores between these 2 independent treatments. The Brier score is the mean squared error (MSE) of the confidence scores (see Eq. \ref{brier_equation}). Thus, we apply a paired t-test 
to determine whether the 
Brier scores are significantly different \cite{ott2015introduction}. We adjust produced p-values using the Benjamini-Hochberg procedure to reduce false discovery risk~\cite{scipy_fdc}.

\section{Result}
\subsection{RQ1 Calibration: LLM Confidence vs. Summary Correctness}
\label{rq1}

\begin{figure}[h]
  \centering
  \begin{subfigure}[b]{0.28\linewidth}
    \centering
    \includegraphics[width=\linewidth]{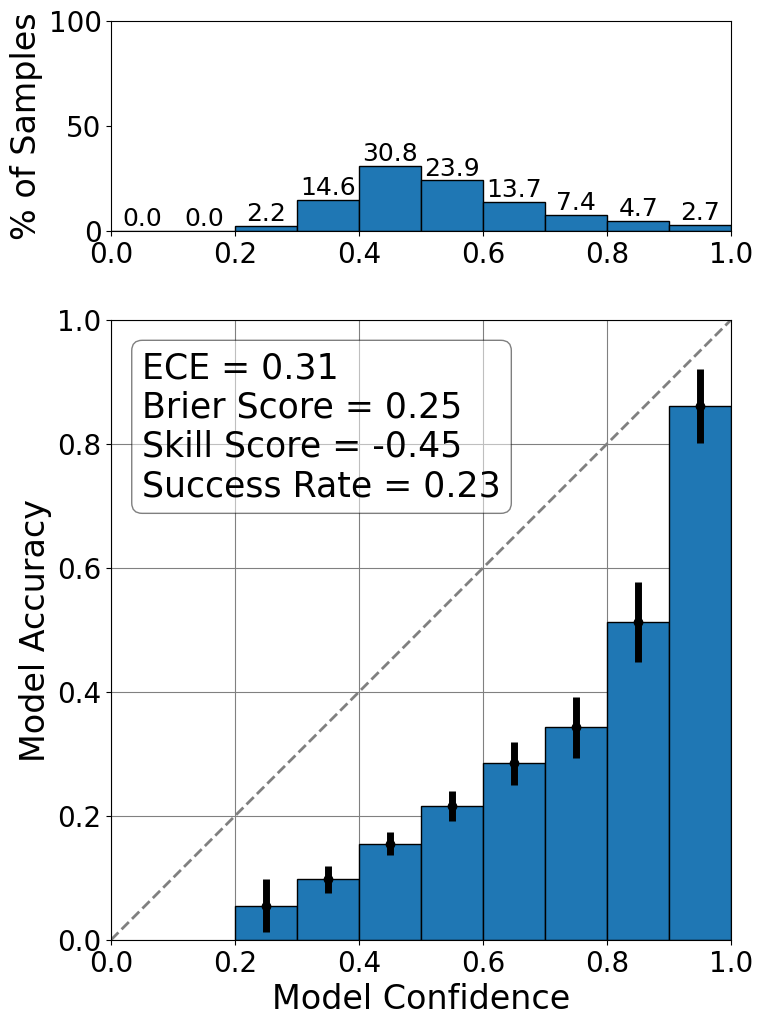}
    \caption{Code Llama 70b}
    \label{codellama_raw_calib}
  \end{subfigure}
  \hspace{10pt}
  \begin{subfigure}[b]{0.275\linewidth} 
    \centering
    \includegraphics[width=\linewidth]{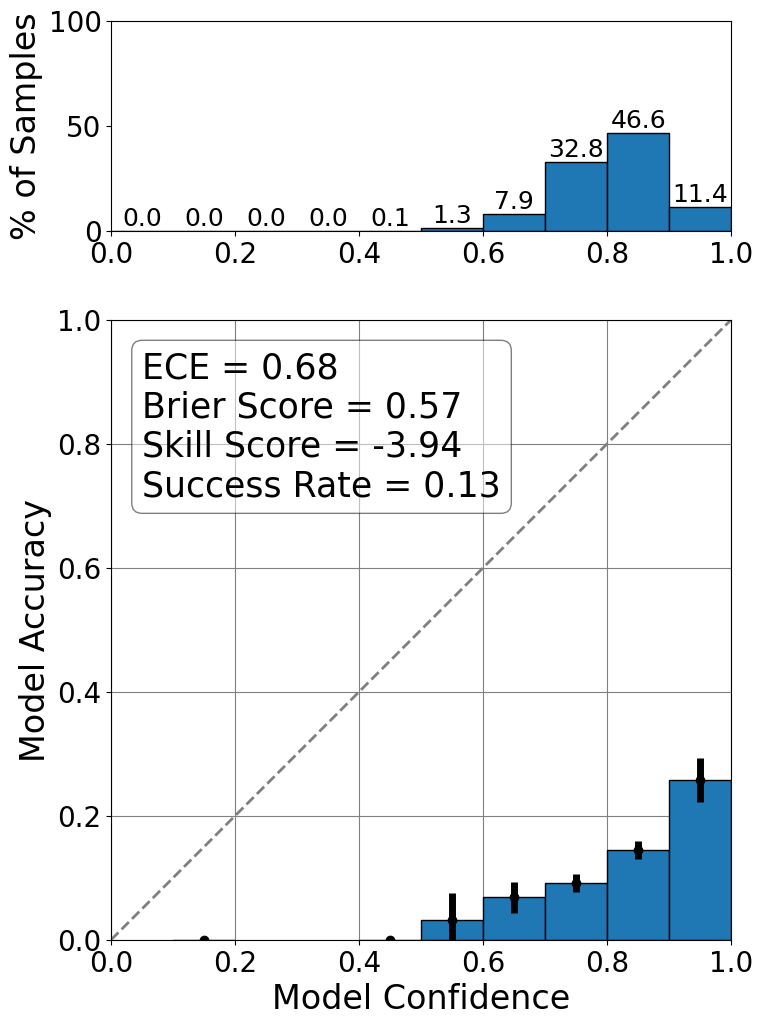} 
    \caption{GPT-3.5 Turbo}
    \label{gpt3.5_raw_calib}
  \end{subfigure}
  \caption{Reliability Plots for models using Retrieval Augmented Few-Shot Summarization on Python methods. The top plots show the frequency distribution of confidence values; the bottom is a reliability diagram.}
  \vspace{-1mm}
  \label{fig:reliability_plots}
\end{figure}

We now present calibration results for summaries generated using few-shots retrieved by BM25. 
Table~\ref{raw_cal}, Figure~\ref{codellama_raw_calib}, and~\ref{gpt3.5_raw_calib} show calibration results with respect to the optimal F1 thresholded BERTScore metric. Both the Brier score and ECE are undesirably high. Additionally, we observed a negative skill score for all cases. The Brier score ranges from 0.25 to 0.57, which indicate worse than random performance. While these calibration values are poor, the reliablity diagrams clearly show some signal. 

We additionally note that model confidences show weak to moderate correlation to BERTScore. In the best case, Codellama-70b Python summaries produced a 0.45 Spearman correlation coefficient \cite{hollander2013nonparametric}. This result is expected since correlation is distinct from calibration. When viewing confidence as a probability indicating the likelihood of binary correctness, 90\% confidence should not indicate 90\% BERTScore; it should indicate \emph{the likelihood} a generated summary is acceptable.



\begin{table}[h]
\centering
\scalebox{0.7}{

\begin{tabular}{llrrrr}
\toprule
 &  & Success Rate & ECE ($\downarrow$) & Brier ($\downarrow$) & Skill score ($\uparrow$) \\
Language & Model &  &  &  &  \\
\midrule
\multirow[t]{3}{*}{Java} & CodeLlama-70b-hf & 0.26 & 0.31 & 0.27 & -0.40 \\
 & deepseek-coder-33b-instruct & 0.23 & 0.34 & 0.27 & -0.53 \\
 & gpt-3.5-turbo & 0.19 & 0.49 & 0.39 & -1.52\\[0.25ex]
\cline{1-6}
\\[-1.75ex]
\multirow[t]{3}{*}{Python} & CodeLlama-70b-hf & 0.23 & 0.31 & 0.25 & -0.45 \\
 & deepseek-coder-33b-instruct & 0.20 & 0.34 & 0.26 & -0.65 \\
 & gpt-3.5-turbo & 0.13 & 0.68 & 0.57 & -3.94 \\
\bottomrule
\end{tabular}
}
\caption{Raw Calibration Metrics with summaries produced by Retrieval Augmented Few-Shot Summarization. ``Success rate" indicates the fraction of output above similarity threshold. Arrows indicate direction of value improvement ($\uparrow$ means higher is better, $\downarrow$ the reverse)}
\label{raw_cal}
\centering
\vspace{-0.5cm}
\end{table}

\subsection{RQ2: Platt Rescaled vs. Raw Calibration}

\label{sec:rescaling}
\begin{figure}[h]
  \centering
  \begin{subfigure}[b]{0.28\linewidth}
    \centering
    \includegraphics[width=\linewidth]{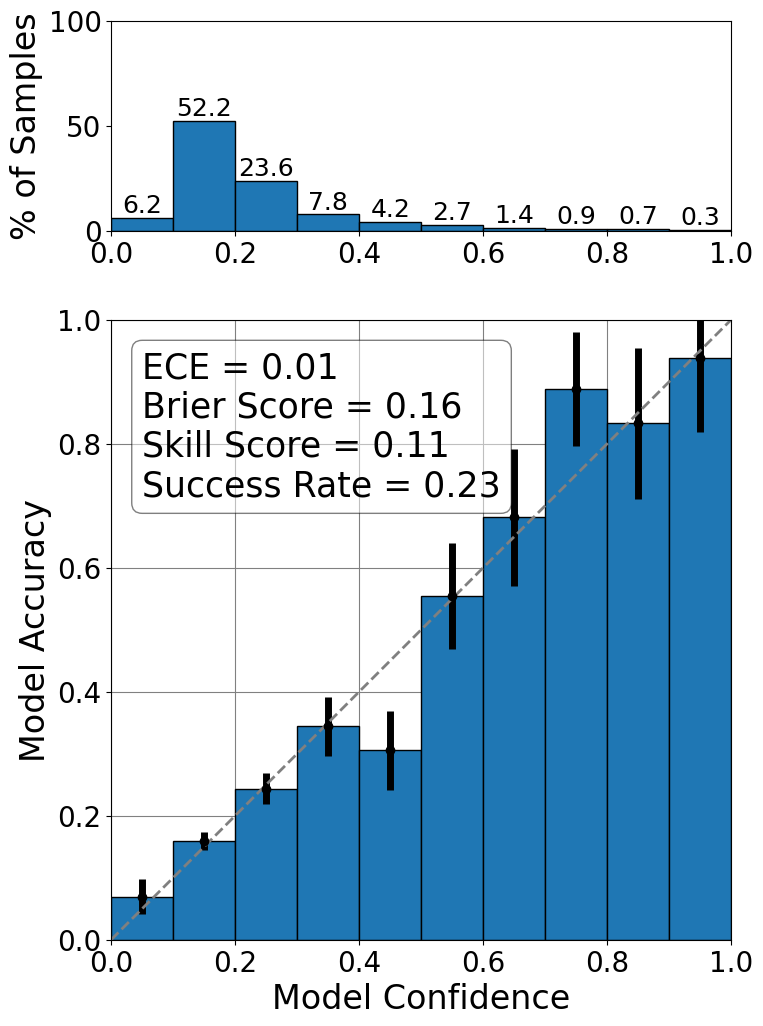}
    \caption{Code Llama 70b}
    \label{fig:figure1}
  \end{subfigure}
  \hspace{10pt}
  \begin{subfigure}[b]{0.28\linewidth} 
    \centering
    \includegraphics[width=\linewidth]{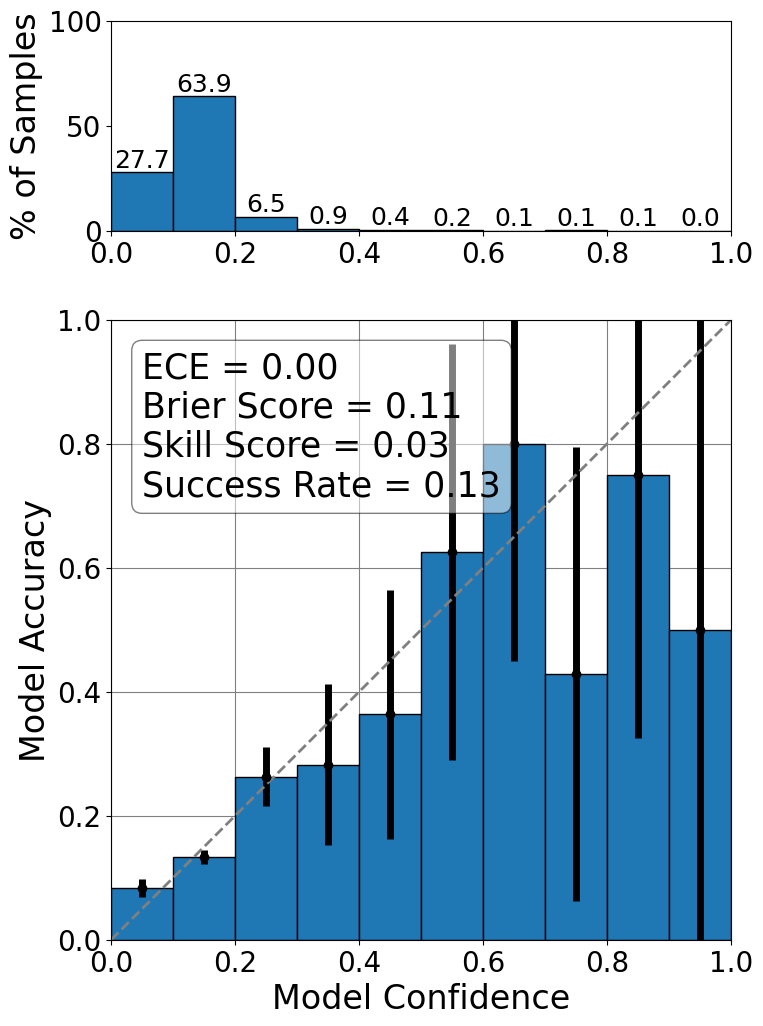} 
    \caption{GPT-3.5 Turbo}
    \label{fig:figure2}
  \end{subfigure}
  \caption{Rescaled Reliability Plots for models using Retrieval Augmented Few-Shot for summarization of Python Methods. Compared to Fig. \ref{fig:reliability_plots}, the rescaled blue bars are closer to the diagonal, and the distribution of confidence values changes (top plots) as they are scaled.}
  \label{fig:rescaled_reliability_plots}
\end{figure}

We now report how calibration changes with rescaling. Table~\ref{platt_calib} and Figure~\ref{fig:rescaled_reliability_plots} show significantly improved calibration for every configuration of model and programming language. All skill scores are positive, and both DeepSeek-Coder-33B and CodeLlama-70b exhibit very high skill scores (0.11) for Java and Python respectively. The Brier scores now range from 0.11 to 0.17 and the ECE ranges from 0.01 to 0.03, indicating good calibration. Although the skill score for GPT-3.5-Turbo is improved after rescaling, it is much lower than the other models. 
To summarize, the raw mean token probability is not well-calibrated, but the rescaled mean token probability is well-calibrated --- with respect to thresholded BERTScore. 

\begin{table}[h]
\centering
\scalebox{0.7}{

\begin{tabular}{llrrrr}
\toprule
Language &                       Model &  Success Rate &  ECE ($\downarrow$) &  Brier ($\downarrow$) &  Skill score ($\uparrow$) \\
\midrule
    Java &               CodeLlama-70b &          0.26 &                0.03 &                  0.17 &                      0.10 \\
         & deepseek-coder-33b-instruct &          0.23 &                0.02 &                  0.16 &                      0.11 \\
         &               gpt-3.5-turbo &          0.19 &                0.03 &                  0.15 &                      0.05 \\
  Python &               CodeLlama-70b &          0.23 &                0.01 &                  0.16 &                      0.11 \\
         & deepseek-coder-33b-instruct &          0.20 &                0.01 &                  0.15 &                      0.08 \\
         &               gpt-3.5-turbo &          0.13 &                0.01 &                  0.11 &                      0.03 \\
\bottomrule
\end{tabular}
}
\caption{Platt Rescaled Calibration using Retrieval Augmented
Few-Shot Summarization on Java methods}
\label{platt_calib}
\centering
\vspace{-0.8cm}
\end{table}

\begin{table}[h]
\centering
\scalebox{0.7}{

\begin{tabular}{llrrrr}
\toprule
Language &                       Model &  Success Rate &  ECE ($\downarrow$) &  Brier ($\downarrow$) &  Skill score ($\uparrow$) \\
\midrule
    Java &               CodeLlama-70b &          0.28 &                0.02 &                  0.19 &                      0.08 \\
         & deepseek-coder-33b-instruct &          0.24 &                0.02 &                  0.16 &                      0.11 \\
         &               gpt-3.5-turbo &          0.24 &                0.04 &                  0.17 &                      0.06 \\
  Python &               CodeLlama-70b &          0.26 &                0.02 &                  0.16 &                      0.19 \\
         & deepseek-coder-33b-instruct &          0.23 &                0.02 &                  0.16 &                      0.11 \\
         &               gpt-3.5-turbo &          0.14 &                0.01 &                  0.12 &                      0.01 \\
\bottomrule
\end{tabular}
}
\caption{Platt Rescaled Calibration using Automatic Semantic Augmentation + Retrieval Augmented Few-Shot for summarization of Java methods}
\label{asap}
\centering
\vspace{-0.5cm}
\end{table}

\vspace{0.05in}
\noindent\textbf{Calibration changes with ASAP.}
Table~\ref{asap} presents the results comparing calibration applying ASAP. We observe that, augmented with static analysis products, every model achieves a higher success rate for both Python and Java. For Java, the skill score decreases with ASAP, whereas for Python, the skill score increases for CodeLlama-70B and deepseek-coder-33b-instruct. It's noteworthy that Java is a strictly typed and verbose language; the additional information
provided by ASAP in the prompts may be redundant, and distort the model's confidence. However, for Python, ASAP may provide necessary information that might be missing in the function body itself (e.g., variable type). Incorporating this information may be helping improve both success rate and calibration. Hence, the trade-off exists, at least for Java, where transitioning from BM25 to ASAP yields better performance but worst calibration.

\vspace{.1cm}

\begin{table*}
\centering
\scalebox{0.7}{

\begin{tabular}{llrrrrrrrr}
\toprule
 &  & \multicolumn{2}{c}{Success Rate} & \multicolumn{2}{c}{ECE ($\downarrow$)} & \multicolumn{2}{c}{Brier ($\downarrow$)} & \multicolumn{2}{c}{Skill score ($\uparrow$)} \\
 \cmidrule(lr){3-4}
 \cmidrule(lr){5-6}
 \cmidrule(lr){7-8}
 \cmidrule(lr){9-10}
 & Threshold Type & High P & High R & High P & High R & High P & High R & High P & High R \\
Language & Model &  &  &  &  &  &  &  &  \\
\midrule
\multirow[t]{3}{*}{Java} & CodeLlama-70b & 0.06 & 0.25 & 0.00 & 0.02 & 0.04 & 0.17 & 0.22 & 0.08 \\
 & deepseek-coder-33b-instruct & 0.04 & 0.22 & 0.01 & 0.01 & 0.02 & 0.15 & 0.35 & 0.12 \\
 & gpt-3.5-turbo & 0.02 & 0.20 & 0.01 & 0.05 & 0.02 & 0.15 & 0.02 & 0.08\\ [0.5ex]
\cline{1-10} 
\\[-1.5ex]
\multirow[t]{3}{*}{Python} & CodeLlama-70b & 0.05 & 0.24 & 0.01 & 0.02 & 0.04 & 0.15 & 0.27 & 0.18 \\
 & deepseek-coder-33b-instruct & 0.01 & 0.20 & 0.02 & 0.01 & 0.02 & 0.14 & 0.28 & 0.10 \\
 & gpt-3.5-turbo & 0.00 & 0.10 & 0.00 & 0.00 & 0.00 & 0.10 & -0.00 & 0.01 \\
\bottomrule
\end{tabular}
}
\caption{Calibration under high precision and recall thresholds. Results are Platt Rescaled and summaries are generated using Automatic Semantic Augmentation + Retrieval Augmented Few-Shot for summarization.}
\vspace{-0.7cm}
\label{thresholds_calibration}
\end{table*}

\vspace{.1cm}

\noindent\textbf{Calibration effects of thresholding}
So far, we have evaluated calibration with respect to a threshold and metric which offers a balance between very strict and lenient classifications (which has the optimal F1-score). How about calibration with respect to a high precision and high recall measures of correctness? Table \ref{thresholds_calibration} presents calibration results for thresholded metrics with high precision and recall respectively. 

As expected, when using a high precision correctness measure (BERTScore thresholded at 0.89), the model's success rate is very low – virtually 0 for GPT-3.5-Turbo summaries of Python methods. For both Python and Java, every model except GPT-3.5-Turbo has excellent calibration with skill scores over 0.2. This high skill is partly due to the very low success rate (see equation \ref{skill_score_eq}). However, in addition to correctly producing low confidence for the many incorrect summaries, the Platt rescaler also learns to produce higher confidence for the few correct ones.

When using a high recall correctness measure (SentenceBERT thresholded at 0.80), the success rate is significantly higher. We find similar calibration results to Table \ref{asap}. Besides GPT-3.5-Turbo on Python methods, skill scores are over 0.08, indicating good calibration. Thus, 
even though more summaries are accepted as being above the threshold, the model's output confidence (after rescaling) provides a reliable indication as to whether the summary might adequately resemble a human-produced summary.

Thus, we find the choice of threshold has a substantial effect on calibration, and so their usage should be managed with care.

\subsection{RQ3: Effect of Token Position on Confidence}
\label{rq3_results}
\begin{figure}[h!]
  \centering
    \includegraphics[width=0.75\linewidth]{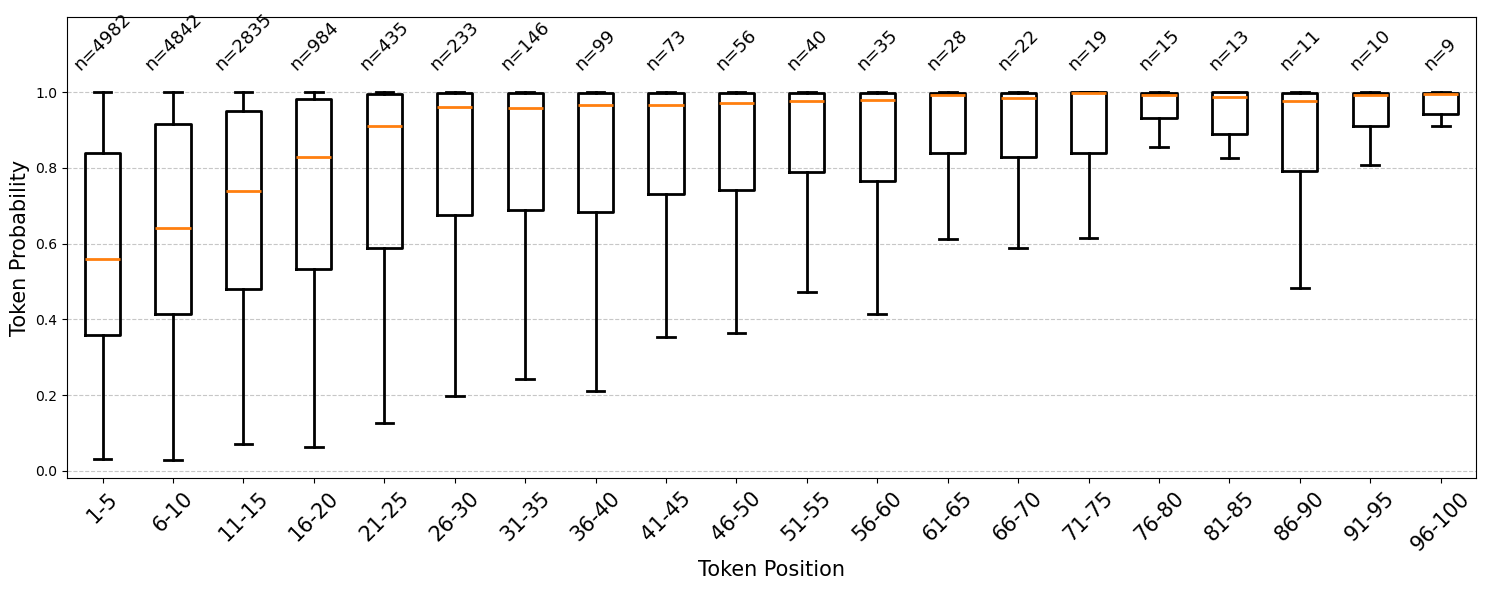}
    \caption{Distributions of token probabilities by their position in a generated summary. Summaries are of Java methods by CodeLlama-70b using Retrieval-Augmented Few-Shot prompting.}
    \label{token_pos_vs_prob}
\end{figure}

We now present the relationship between token position and model confidence and its effect on calibration. Figure \ref{token_pos_vs_prob} shows that, as a token's position in a Java code summary increases, CodeLlama-70b's confidence in the token also tends to increase. After 20 tokens, the median token probability for each distribution plateaus to over 0.9. Similar trends (where tokens late in the sequence exhibit high probabilities) are seen for every configuration of model, programming language, and prompting method. However, in some cases, this relationship is exhibited to a lesser extent e.g. in GPT-3.5-Turbo-generated Python summaries, the median of token probabilities at positions 1-5 is already over 0.8. Although later tokens still tend to have even higher probabilities, excluding them during the mean probability calculation has a smaller effect since the mean probability of the cutoff sequence is already very large.

\begin{figure}[h!]
  \centering
  \begin{subfigure}[t]{0.35\linewidth}
    \centering
    \includegraphics[width=\linewidth]{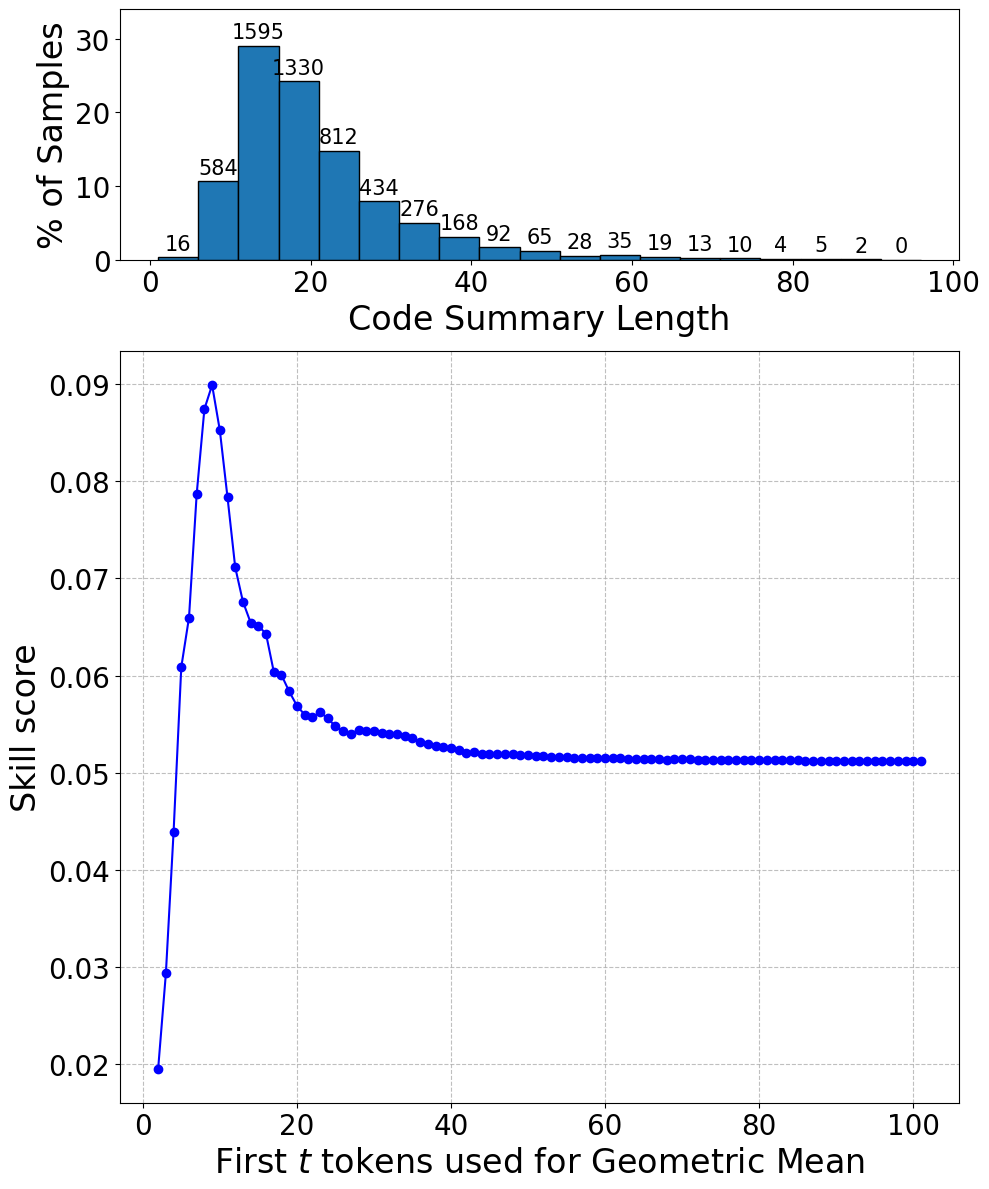} 
    \caption{GPT-3.5 Turbo on Java methods using retrieval-augmented prompting. \emph{Summarization Success Rate}: 0.19}
    \label{fig:figure1}
  \end{subfigure}
  \hspace{10pt}
  \begin{subfigure}[t]{0.35\linewidth} 
    \centering
    \includegraphics[width=\linewidth]{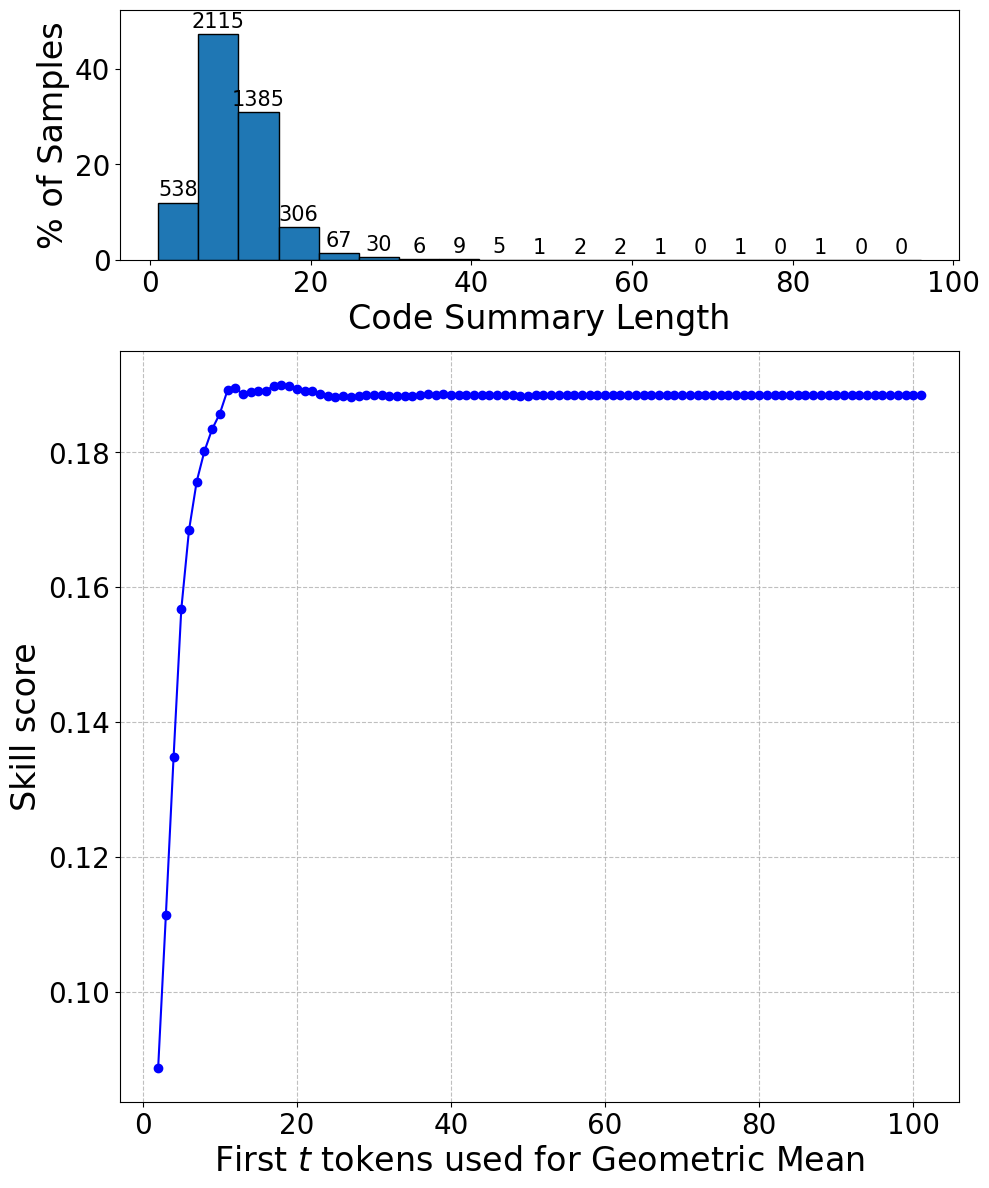}
    \caption{CodeLlama-70b on Python methods using Automatic Semantic Augmentation. \emph{Summarization Success Rate}: 0.26}
    \label{fig:figure2}
  \end{subfigure}
  \vspace{-0.2cm}
  \caption{The bottom plots show the skill score achieved, after platt scaling, when taking the arithmetic mean of the first \emph{t} tokens. The top plots show the frequency distribution of generated code summary lengths. In (b), since there are very few generated summaries longer than 20 tokens, only using early tokens when computing confidences has little overall effect on Skill Score.}
  \label{fig:token_cutoff_vs_ss}
\end{figure}

Figure \ref{fig:token_cutoff_vs_ss} shows the skill score achieved when only including the first $t$ tokens in the geometric mean. In Figure \ref{fig:figure1}, all first 7 tokens carry significant signal. However, including any more tokens reduces skill score since their probabilities are very large, smoothing the total mean towards high values which are not representative of uncertainty in the summary. In Figure \ref{fig:figure2}, only including early tokens in the geometric mean has little effect, primarily due to CodeLlama-70b generating short code summaries for Python (see the top plot in Figure \ref{fig:figure2}). Since the proportion of generated code summaries with over 15 tokens is only 9.41\%, computing a mean using the first \emph{e.g.} 15 tokens has a small effect on skill score (which is measured across all summaries).

Following Section \ref{token_cutoff}, by using just the first \emph{t}, instead of all tokens in computing the geometric mean, we find a statistically significant ($\alpha = 0.01$) reduction in the Brier Scores (after Benjamini-Hochberg correction) for some but not all configurations of model, programming language, and prompting method. For Python summaries, we only find statistically significant reductions for CodeLlama-70b and DeepSeek-Coder-33b with few-shot prompting. For Java summaries, we find significant reductions for all models and prompting methods except DeepSeek-Coder-33b few-shot prompted.

Why is there a significant reduction in Brier Score in some cases and not others? Some configurations generate shorter summaries \emph{e.g.} Figure \ref{fig:figure2}. We find that code summaries generated for Python methods tend to be shorter than code summaries for Java. Additionally, the selected hyperparameter $t$ might be poor; for Java summaries generated with CodeLlama-70b and ASAP prompting, some folds use $t$=4 and may thus lose significant information. Finally, probabilities early in the summary may already be large \emph{e.g.}  GPT-3.5-Turbo-generated Python summaries discussed earlier in the section.

In summary, our analysis suggests that the ideal practical configuration of prompting strategy (ASAP and/or RAG) and token length (for confidence) depends on the setting, and could be empirically selected. 

\section{Discussion: Threats \& Limitations}
\label{discussion}

\subsection{Do LLMs surpass code summaries found in open-source projects?}
Our study compares LLM-generated code summaries against human-written ones collected from open-source projects. If LLMs surpass the quality of open-source code summaries, using them as references is undesirable. Instead, curated code summaries or reference-free evaluation techniques would be necessary. However, even as language model performance improves, calibration remains important; this paper clarifies, and lays out methodologies for studying calibration, even under future improved summarization and evaluation methods. It should be noted, however, that it's not yet clear to what degree LLM-generated summaries surpass open-source ones in helping developers. Existing work suggests LLMs may surpass humans in writing code summaries \cite{sun2024sourcecodesummarizationera, su2024contextawarecodesummarygeneration}. \cite{su2024distillgptcodesummarization} find that participants generally preferred GPT-3.5 generated code summaries over open-source ones, often rating them higher in terms of completeness and accuracy. However, LLM-generated summaries and LLM evaluations have not been explicitly judged based on on-task programming performance; the evidence that LLM-generated summaries are better than human summaries relies on human ratings, rather than help with on-task performance; \cite{stapleton2020human} find that human ratings are uncorrelated with performance on comprehension questions, thus casting doubt on human ratings of completeness and accuracy with respect to code. It's important to note here that the human ratings we use are for the simpler purpose of selecting a threshold to evaluate the similarity to available human-written summaries (for our subsequent experiments), not to evaluate completeness and accuracy with respect to the code. Our use does not require the human ratings to be based on the same kind of working knowledge of the codebase and software process that the original human summary writer would have possessed.

\subsection{Limitation: Summary Intent Capture}
Human developers have different intents (how/what/why) when writing summaries~\cite{hu2022practitioners,geng2024large}, so a user might request summaries to be generated for a particular intent \cite{su2024contextawarecodesummarygeneration}. However, these different intents are not explicitly captured by similarity metrics. A reliable way of integrating intent into automated evaluation metrics and confidence scores remains for future work. 

\label{api}


\subsection{Limitation: Platt scaling}
Platt scaling is a popular and widely used rescaling strategy. However, another very popular rescaling technique is temperature scaling, which is found to be better than Platt scaling in certain scenarios. In this paper, we could not use temperature scaling because it requires access to complete softmax layers, which are unavailable from the OpenAI\footnote{https://openai.com} and TogetherAI\footnote{https://www.together.ai} APIs. In the future, we hope to incorporate temperature when we have access to the complete softmax layer.


\subsection{Self-Reflection Calibration}
By ``reflective," we mean the ability of an instruction-tuned model to evaluate its own output. We used two reflective measures to compute model confidence: a logit-based one and a verbalized one. 

For the logit-based measure, we presented the LLM with the code and model-generated summary, then asked it whether the summary resembled something a human might write by generating a True/False response. Our prompts were adapted from Spiess et al.~\cite{spiess2024calibration}. Using API access to the top-5 token probabilities, we extracted the logit corresponding to the token True as the model's confidence score.

The logit-based measure was only feasible with GPT-3.5-Turbo. Other models, like CodeLlama-70B, failed to generate valid Boolean outputs. Additionally, due to API limitations, we didn't have access to the top-5 log-probabilities for models other than GPT-3.5-Turbo. GPT-3.5-Turbo exhibited severe overconfidence: 94\% of predictions had confidence scores above 0.9 and its performance yielded a negative skill score. Rescaling the logits transformed them to near the unskilled base rate.

For the verbalized measure, we asked models to rate the summary's resemblance to what a human might write on a scale from 1 to 4. We evaluated both zero-shot and few-shot prompting; our prompts followed the human evaluation framework of ~\cite{haque2022semantic}. In the zero-shot setting, CodeLlama-70B failed to produce valid scores, but it succeeded under few-shot prompting using 4 examples from the Haque dataset. Still, in both prompting setups, model outputs were poorly calibrated. We experimented with higher temperatures and majority voting across multiple samples, but these techniques did not improve calibration.

Therefore, we can conclude that the models are not well-calibrated for any of the self-reflective measures studied.

\section{Related Work}
LLMs are widely used in many software engineering tasks~\cite{fan2023large,hou2023large}, including code summarization~\cite{ahmed2022few,ahmed2024automatic}. Though LLMs are drawing attention from the community, it is well-known that they generate more buggy text/code than fixed text/code~\cite{jesse2023large}. Consequently, the reliability of the model-generated output remains very low. To address this reliability issue, researchers have started looking into calibration in Software Engineering~\cite{zhou2024calibration,zhang2023pace}. 

Calibration is a well-studied domain in NLP~\cite{guo2017calibration}, primarily for classification problems where model log-probabilities are used as a measure of model confidence. Considerable work in the generative setting has also been done in NLP \cite{malinin2021uncertainty}, \cite{kuhn2023semantic}, \cite{ott2018}, \cite{wang-etal-2020-inference}. Though calibration is also being studied in the SE community~\cite{zhou2024calibration,zhang2023pace, spiess2024calibration}, to the best of our knowledge, there is no study on calibration for generative models for code summarization. Compared to existing work in NLP, the notion of "correctness" for code summaries is distinct and multi-faceted, as discussed in \ref{sec:summaryeval}. Compared to work in program repair or synthesis, measures of correctness for code summaries are continuous. To address these aspects, we explored thresholding similarity measures. Our results regarding the effect of programming language choice, and using static analysis products in prompt, on calibration for code summaries, are also SE-specific. Finally, to our knowledge, the observation that uncertainty is better represented by earlier tokens than later ones is a novel contribution in a SE context.

SIDE is another reference-free evaluation metric for code summaries, aiming to measure a summary's content adequacy \cite{mastropaolo2024evaluating}. However, as Mastorpaolo et al. discuss, it is limited in capturing content adequacy e.g. giving high scores to relevant but generic summaries, and should be viewed as complementary to similarity metrics. Our method accurately captures our measure of correctness as shown, yet also has limitations arising from using similarity as the metric. As discussed, in section \ref{sec:summaryeval}, both SIDE and a calibrated confidence score can be presented to the user since they can indicate different aspects of summary quality. 

In SE, LLMs are increasingly popular, but questions about reliability persist. We anticipate more studies on calibration for different problems in the near future to enhance the reliability of model output.

\section{Conclusion \& Acknowledgements}
Code summaries are valuable to maintainers, and LLMs can generate good summaries from code. However, LLMs generate summaries that vary in quality: sometimes they resemble human-written summaries, and sometimes not. This indicates a need for developers to know when a summary might be \emph{sufficiently similar} to a human-written summary. This is a calibration problem: can the model provide a reliable indication of its confidence that a generated summary is ``good enough"? We address this problem by selecting a metric known from the literature to be a good proxy for similarity to human-produced summaries, and then thresholding this metric to get an indication of sufficient similarity. We then evaluate the calibration
of a range of prompting approaches, together with Platt scaling. We find that
in some cases, the first few tokens may provide the best calibration; we also
find both retrieval-augmented few-shot prompting and ASAP can be advantageous. We also find that reflection-based
approaches do not provide better calibration. To our knowledge, we are the first
to evaluate whether LLM-based approaches to code summarization are well calibrated; our work
contributes both a methodology, and several useful approaches, to this important question. 

We gratefully acknowledge support from NSF CISE, Grant Number 2107592, entitled: 
\emph{``SHF:Medium: Studying and Exploiting the Bimodality of Software"}.

\section{Data Availability}

Our code and datasets are made available on \href{https://github.com/yuvrajvirk/CalibrationLLMsCodeSummaries/}{GitHub}.

\bibliographystyle{ACM-Reference-Format}


\end{document}